

\font\titlefont = cmr10 scaled\magstep 4
\font\sectionfont = cmr10

\font\teenyfont = cmr5

\magnification = 1200

\global\baselineskip = 1.2\baselineskip
\global\parskip = 4pt plus 0.3pt
\global\nulldelimiterspace = 0pt

\predisplaypenalty 1000


\def\endignore{}
\def\ignore #1\endignore{}

\newcount\dflag
\dflag = 0


\def\monthname{\ifcase\month
\or Jan \or Feb \or Mar \or Apr \or May \or June%
\or July \or Aug \or Sept \or Oct \or Nov \or Dec
\fi}




\def\endid{}
\def\id#1\endid{\number\day\ \monthname \number\year
\hfill #1}

\def\endtitle{}
\def\title#1\endtitle{\vskip.3in\titlefont
\global\baselineskip = 2\baselineskip
#1\vskip.4in
\baselineskip = 0.5\baselineskip\rm}

\def\lblfoot{This work was supported by the Director, Office of Energy
Research, Office of High Energy and Nuclear Physics, Division of High
Energy Physics of the U.S. Department of Energy under Contract
DE-AC03-76SF00098.}

\def\endauthors{}
\def\authors#1\endauthors{
#1\if\dflag = 0
\footnote{}{\noindent\lblfoot}\fi}

\def\endabstract{}
\def\abstract#1\endabstract{\vskip .3in%
\centerline{\sectionfont\bf Abstract}%
\vskip .1in%
\noindent#1%
\ifnum\dflag = 0
\footline = {\hfil}\pageno = 0
\vfill\eject
\pageno = 1\footline{\centerline{\sectionfont\folio}}
\fi\ifnum\dflag = 2
\footline = {\hfil}\pageno = 0
\vfill\eject
\fi}

\newcount\nsection
\newcount\nsubsection

\def\section#1{\global\advance\nsection by 1
\global\nsubsection = 0
\bigskip\noindent\sectionfont \bf \number\nsection.\ #1
\nobreak\medskip\rm\nobreak}

\def\subsection#1{\global\advance\nsubsection by 1
\bigskip\noindent\sectionfont \it \number\nsection.\number\nsubsection.\ #1%
\nobreak\medskip\rm\nobreak}

\def\appendix#1#2{\bigskip\noindent%
\sectionfont \bf Appendix #1.\ #2
\nobreak\medskip\rm\nobreak}


\newcount\nref
\global\nref = 1

\def\ref#1#2{\xdef #1{[\number\nref]}
\ifnum\nref = 1\global\xdef\therefs{\noindent[\number\nref] #2\ }
\else
\global\xdef\oldrefs{\therefs}
\global\xdef\therefs{\oldrefs\vskip.1in\noindent[\number\nref] #2\ }%
\fi%
\global\advance\nref by 1
}

\def\listrefs{\vfill\eject\section{References}\therefs}


\newcount\nfig
\global\nfig = 1

\def\fg#1\efig{\vskip .5in\noindent Fig.\ \number\nfig:\ #1%
\global\advance\nfig by 1}


\newcount\cflag
\newcount\nequation
\global\nequation = 1
\def\eqlabel{(1)}

\def\nexteqno{\ifnum\cflag = 0
\global\advance\nequation by 1
\fi
\global\cflag = 0
\xdef\eqlabel{(\number\nequation)}}

\def\lasteqno{\global\advance\nequation by -1
\xdef\eqlabel{(\number\nequation)}}

\def\label#1{\xdef #1{(\number\nequation)}
\ifnum\dflag = 1
{\escapechar = -1
\xdef\draftname{\teenyfont\string#1}}
\fi}

\def\clabel#1#2{\xdef\eqlabel{(\number\nequation #2)}
\global\cflag = 1
\xdef #1{\eqlabel}
\ifnum\dflag = 1
{\escapechar = -1
\xdef\draftname{\string#1}}
\fi}

\def\cclabel#1#2{\xdef\eqlabel{#2)}
\global\cflag = 1
\xdef #1{\eqlabel}
\ifnum\dflag = 1
{\escapechar = -1
\xdef\draftname{\string#1}}
\fi}


\def\eeq{}

\def\eqnn #1\eeq{$$ #1 $$}

\def\eq #1\eeq{\xdef\draftname{\ }
$$ #1
\eqno{\eqlabel \rlap{\ \draftname}} $$
\nexteqno}



\def\eol{& \eqlabel \rlap{\ \draftname} \crcr
\nexteqno
\xdef\draftname{\ }}

\def\eeol{& \eqlabel \rlap{\ \draftname}
\nexteqno
\xdef\draftname{\ }}



\def\eqa #1\eeq{\xdef\draftname{\ }
$$ \eqalignno{ #1 } $$
\global\cflag = 0}



\def\myinstitution{
    \centerline{\it Theoretical Physics Group}
    \centerline{\it Lawrence Berkeley Laboratory}
    \centerline{\it 1 Cyclotron Road}
    \centerline{\it Berkeley, California 94720}
}


\def\jref#1#2#3#4{{\it #1} {\bf #2}, #3 (#4)}

\def\NC#1#2#3{\jref{Nuovo Cim.}{#1}{#2}{#3}}
\def\NPB#1#2#3{\jref{Nucl.\ Phys.}{B#1}{#2}{#3}}
\def\PA#1#2#3{\jref{Physica}{#1A}{#2}{#3}}
\def\PLB#1#2#3{\jref{Phys.\ Lett.}{#1B}{#2}{#3}}

\def\PRD#1#2#3{\jref{Phys.\ Rev.}{D#1}{#2}{#3}}

\def\PRL#1#2#3{\jref{Phys.\ Rev.\ Lett.}{#1}{#2}{#3}}
\def\PRV#1#2#3{\jref{Phys.\ Rev.}{#1}{#2}{#3}}


\def\goto{\mathop{\rightarrow}}
\def\gotoo{\mathop{\longrightarrow}}
\def\mapstoo{\mathop{\longmapsto}}


\def\myint{\int\mkern-5mu}
\def\frac#1#2{{{#1} \over {#2}}\,}  


\def\Dsl{\hbox{/\kern-.6000em\rm D}} 



\def\mybar#1{\kern 0.8pt\overline{\kern -0.8pt#1\kern -0.8pt}\kern 0.8pt}
\def\sla#1{\raise.15ex\hbox{$/$}\kern-.57em #1}
\def\Sla#1{\kern.15em\raise.15ex\hbox{$/$}\kern-.72em #1}

\def\roughly#1{\mathrel{\raise.3ex\hbox{$#1$\kern-.75em%
    \lower1ex\hbox{$\sim$}}}}

\def\gsim{\roughly>}

\def\scr#1{{\cal #1}}






\def\tr{\mathop{\rm tr}}


\def\bra#1{\langle #1 |}
\def\ket#1{| #1 \rangle}

\def\avg#1{\langle #1 \rangle}


\def\ddp#1#2{\frac{d^{#1}#2}{(2\pi)^{#1}}\,}


\def\GIM{Glashow--Iliopoulos--Maiani}

\def\hc{{\rm h.c.}}


\def\GUC{$SU(N)_{UC}$}
\def\KKLR{$SU(K + K')_L \times SU(K + K')_R$}
\def\KLR{$SU(K)_L \times SU(K)_R$}

\id
LBL-32299
\endid

\title
\centerline{Vacuum Alignment in}
\centerline{``Composite Technicolor'' Models}
\endtitle

\authors
\centerline{Markus A. Luty}
\footnote{}{\lblfoot}
\vskip .1in
\myinstitution
\endauthors

\abstract
We consider the question of vacuum alignment in the recently proposed
``composite technicolor'' (CTC) models.
In these models, explicit chiral symmetry breaking due to masses of
electroweak-singlet ``ultrafermions'' is communicated to the quarks and
leptons by a chiral condensate produced by strong ``ultracolor'' gauge
interactions.
In order for these models to work, the ultrafermion condensate must
align in a particular way, driven by the competition between ultrafermion
masses and ``flavor'' gauge boson exchange.
We show that for ultrafermion masses large enough to explain the top quark
mass, order-of-magnitude estimates are sufficient to establish that the
required vacuum alignment cannot occur for perturbative values of the flavor
gauge couplings.
If the top quark gets its mass from some other mechanism, a detailed
calculation
is required.
Using spectral function sum rules and a vector-meson saturation approximation
which is known to work well in QCD, we determine the correct vacuum for this
case in a limit where the flavor gauge bosons are weakly coupled and the
number of ultracolors $N$ is large.
We again find that the vacuum required by CTC models does not occur for
perturbative values of the flavor gauge couplings.
We conclude that there is no evidence that the vacuum aligns as required
in CTC models.
\endabstract


\ref\CTSMidea{R.\ S.\ Chivukula and H.\ Georgi, \PLB{188}{99}{1987}.}

\ref\GIMpap{S.\ L.\ Glashow, J.\ Iliopoulos, and L.\ Maiani,
\PRD{2}{1285}{1970}.}

\ref\CTSM{R.\ S.\ Chivukula, H.\ Georgi, and L.\ Randall,
\NPB{292}{93}{1987}.}

\ref\Georgi{H.\ Georgi in T.\ Muta and K.\ Yamawaki {\it eds.\/},
{\it Proceedings of the 1990 Workshop on Strong Coupling Gauge Theories and
Beyond} (World Scientific, Singapore, 1991).}

\ref\Kingvac{N.\ J.\ Evans, S.\ F.\ King, and D.\ A.\ Ross, Southhampton
Preprint SHEP 91/92-11.}

\ref\Kingmod{S.\ F.\ King, \PRD{45}{990}{1992};
Southhampton preprint SHEP 91/92-7.}

\ref\reffpot{J.\ Goldstone, A.\ Salam, and S.\ Weinberg,
\PRV{127}{965}{1962};
G.\ Jona-Lasinio, \NC{34}{1790}{1964};
B.\ W.\ Lee and J.\ Zinn-Justin, \PRD{5}{3121}{1972};
S.\ Coleman and E.\ Weinberg, \PRD{7}{1888}{1973};
R.\ Jackiw, \PRD{9}{1686}{1974}.}

\ref\effL{J.\ Schwinger, \PLB{24}{473}{1967};
S.\ Coleman, J.\ Wess, and B.\ Zumino, \PRV{117}{2239}{1969};
C.\ G.\ Callan, S.\ Coleman, J.\ Wess, and B.\ Zumino, \PRV{117}{2247}{1969};
S.\ Weinberg, \PA{96}{327}{1979}.}

\ref\NDA{H.\ Georgi and A.\ Manohar, \NPB{234}{189}{1984}.}

\ref\vacalign{R.\ Dashen, \PRV{183}{1245}{1969}; \PRD{3}{1879}{1971};
S.\ Weinberg, \PRD{13}{974}{1976}; \PRD{19}{1277}{1979};
M.\ E.\ Peskin, \NPB{175}{197}{1980};
J.\ Preskill, \NPB{177}{21}{1981}.}

\ref\SFSR{S.\ Weinberg, \PRL{18}{507}{1967};
C.\ Bernard, A.\ Duncan, J.\ Lo Secco, and S.\ Weinberg, \PRD{12}{792}{1975}.}

\ref\PT{For a review and references, see M.\ E.\ Peskin and T.\ Takeuchi,
SLAC preprint SLAC-PUB-5618, to appear in {\it Phys. Rev.} {\bf D}.}

\ref\pp{T.\ Das, G.\ S.\ Guralnik, V.\ S.\ Mathur, F.\ E.\ Low, and
J.\ E.\ Young, \PRL{18}{759}{1967}.}


\section{Introduction}

By far the most difficult problem facing technicolor theories is how to
generate fermion masses and mixings without at the same time giving rise to
flavor-changing neutral currents at unacceptable levels.
In ref.\ \CTSMidea, it was pointed out that a symmetry which guarantees
a \GIM\ (GIM) mechanism \GIMpap\ could in principle
be imposed on the low-energy effective theory arising from a technicolor model.
In ref.\ \CTSM, the generic problems in implementing this idea in models
without elementary scalars were identified, and ``composite technicolor''
(CTC) models were proposed which are supposed to solve these problems.
(See also ref.\ \Georgi.)
These models rely crucially on the assumption that the vacuum aligns in a
particular way, driven by the competition between ultrafermion mass terms and
``flavor'' gauge boson exchange.
In the desired vacuum, the chiral symmetry breaking due to ultrafermion masses
is communicated to the ordinary quarks through massive flavor gauge boson
exchange.
It has been conjectured \Georgi\Kingvac\ that the desired vacuum is obtained
for flavor couplings that are large, but not large enough so that they
themselves break chiral symmetry.

In this paper, we consider the effective potential for this model.
We show that in the case where the ultrafermion masses are sufficiently large
so that this mechanism can give rise to the top quark mass, order-of-magnitude
estimates are sufficient to establish that the required vacuum alignment
cannot occur for perturbative values of the flavor gauge couplings.

We can also imagine models where this mechanism gives rise to the masses
of the first two generations only, and the ultrafermion masses are small.
In this case, a calculation is required to settle the alignment question.
We compute the effective potential for this case in the limit
that the flavor gauge coupling is weak and the number of ``ultracolors''
$N$ is large.
In this limit, the vacuum alignment depends only on a current--current
correlation function which can be estimated using a vector-meson saturation
approximation which is known to work well in QCD.
We find that the lowest-order perturbative effective potential predicts that
the vacuum required by the CTC models occurs only for flavor gauge couplings
well outside the range of validity of the calculation.
We therefore conclude that there is no evidence that the vacuum aligns
as required by the CTC models.

The plan of this paper is as follows:
In section 2, we will introduce the model whose vacuum alignment
we wish to determine, and discuss the form of the relevant effective
potential.
We argue that the required vacuum cannot occur in models where the top quark
mass is given by the CTC mechanism.
In section 3, we briefly review the formalism for addressing the question
of vacuum alignment in QCD-like theories and compute the effective potential
for the case where only light fermion masses occur through the CTC mechanism.
Section 4 contains our conclusions.

\section{The Model}

The model we will consider has a gauge group
\eq
SU(N)_{UC} \times SU(K)_L \times SU(K)_R,
\eeq
where \GUC\ is a strongly-coupled ``ultracolor'' group and \KLR\ is a
``flavor'' gauge group.
The model contains the fermions
\eq
\eqalign{
Q_L & \sim (1, K, 1) \times N, \cr
Q_R & \sim (1, 1, K) \times N, \cr
\psi_L & \sim (N, K, 1), \cr
\psi_R & \sim (N, 1, K), \cr
\chi_L & \sim (N, 1, 1) \times K', \cr
\chi_R & \sim (N, 1, 1) \times K'. \cr}
\eeq
Note that some of the representations are repeated.
All gauge anomalies are easily seen to cancel.
In addition to the gauge interactions, the $\chi$ fermions are assumed to have
Dirac masses of the form
\eq
\chi_L m_\chi \chi_R^{\rm c} + \hc
\eeq
(We work in a basis where $m_\chi$ is diagonal and positive.)
This model is not intended to be realistic by itself, but it is a
crucial building block for the realistic models of refs.\ \CTSM\Georgi.
(See also ref.\ \Kingmod.)
In order for these realistic models to work, the model above must choose a
particular vacuum.
This vacuum alignment question is the subject of this paper.

Suppose that the \GUC\ gauge coupling becomes strong at a scale $\Lambda_{UC}$.
If we neglect $m_\chi$ and the \KLR\ gauge couplings, the theory undergoes
spontaneous symmetry breaking in the pattern
\eq
SU(K + K')_L \times SU(K + K')_R \gotoo SU(K + K')_{L + R}.
\eeq
The $\chi$ masses and \KLR\ gauge couplings break the symmetry explicitly
down to
\eq
\label\exgroup
SU(K)_L \times SU(K)_R \times \left[ U(1) \right]^{K'}.
\eeq
(Upon diagonalization, the mass matrix $m_\chi$ preserves $K'$ separate
$U(1)$ ``$\chi$-number'' symmetries.)
If $m_\chi \ll \Lambda_{UC}$ and the \KLR\ couplings are small, then we can
treat the explicit breaking effects perturbatively.

When the \GUC\ gauge group becomes strong, it will give rise to the condensate
\eq
\label\cond
\avg{\Psi_{Lj} \Psi^{\rm c}_{Rk}} = U_{jk} \Lambda_{UC}^3,
\eeq
where
\eq
\label\thebasis
\Psi \equiv \pmatrix{\psi \cr \chi \cr},
\eeq
and $U$ is a $SU(K + K')$ matrix.
In the limit where \KKLR\ is an exact symmetry, $U$ has no physical
significance, since we can perform a chiral rotation to set $U = 1$.
However, in the presence of explicit breaking, $U$ determines the orientation
of
the exact symmetry group \exgroup\ inside the unbroken $SU(K + K')_{L + R}$
group.

The physics below the scale $\Lambda_{UC}$ can be summarized by an effective
lagrangian containing the Nambu--Goldstone bosons (NGB's) and the
\KLR\ gauge fields \effL.
This lagrangian should be viewed as a systematic simultaneous expansion in
derivatives and the symmetry-breaking parameters.
The NGB fields $\pi_A$ are collected into a $SU(K + K')$ matrix
\eq
\Sigma = e^{i\pi_A L_A / f}
\eeq
transforming under $SU(K + K')_L \times SU(K + K')_R$ as
\eq
\Sigma \mapstoo U_L \Sigma U_R^\dagger.
\eeq
(Here $L_A$ are $SU(K + K')$ generators normalized so that
$\tr L_A L_B = \delta_{AB}$.)
The gauge covariant derivative acting on $\Sigma$ is
\eq
D_\mu \Sigma = \partial_\mu \Sigma + ig A_{L\mu a} T_{La} \Sigma
- ig A_{R\mu a} \Sigma T_{Ra},
\eeq
where
\eq
T_{La} = T_{Ra} = \pmatrix{t_a & 0 \cr 0 & 0 \cr}
\eeq
in the basis defined by eq.\ \thebasis, and we have set the left- and
right-handed flavor gauge couplings equal for simplicity.
We normalize the gauge generators such that $\tr t_a t_b = \delta_{ab}$.

The effective lagrangian contains the derivative terms
\eq
\scr L = \frac{f^2}{2} \tr \left[
(D^\mu \Sigma)^\dagger D_\mu \Sigma \right] + \cdots.
\eeq
In addition, the theory contains non-derivative terms which give rise to a
tree-level potential term for the NGB's:
\eq
\label\potform
\eqalign{
V(\Sigma) & = a f^3 \tr\left( \Sigma^\dagger M_\chi + \hc \right)
+ O(M_\chi^2) \cr
&\qquad + \frac{b_1 g^4}{16\pi^2} \, f^4
\tr\left( T_{La} \Sigma T_{Rb} \Sigma^\dagger \right)
\tr\left( T_{La} \Sigma T_{Rb} \Sigma^\dagger \right) \cr
&\qquad + \frac{b_2 g^4}{16\pi^2} \, f^4
\tr\left( T_{La} \Sigma T_{Rb} \Sigma^\dagger
T_{La} \Sigma T_{Rb} \Sigma^\dagger \right) + O(g^6), \cr}
\eeq
where
\eq
M_\chi \equiv \pmatrix{0 & 0 \cr 0 & m_\chi \cr}.
\eeq
Based on the ideas of ``naive dimensional analysis'' \NDA, we expect that
$|b_1|, |b_2| \sim 1$.
This is confirmed by the computations of the next section.

The full effective potential includes loop corrections.
We write
\eq
V_{\rm eff}(\Sigma) = V(\Sigma) + \delta V(\Sigma),
\eeq
where $\delta V$ denotes the loop corrections.
These can be computed by standard methods \reffpot.
The contribution from one gauge boson loop gives
\eq
\label\vcorr
\delta V(\Sigma) = \frac{3}{64\pi^2} \tr M^4
\ln\frac{M^2}{\mu^2},
\eeq
where
\eq
M^2 \equiv g^2 f^2 \pmatrix{1 & -t \cr -t & 1 \cr},
\qquad
t_{ab} \equiv \tr\left( T_{La} \Sigma T_{Rb} \Sigma^\dagger \right).
\eeq
$M^2$ is the tree-level mass matrix of the \KLR\ gauge bosons in the
background field described by $\Sigma$, and $\mu$ is an arbitrary
renormalization scale.
Eq.\ \vcorr\ is the leading correction for small $g$.
Note that the nonanalytic correction term are {\it \'a priori} of
the same order of magnitude as the tree-level term.

Minimizing the full effective potential gives a vacuum expectation value for
the NGB fields which specifies the alignment of the vacuum.
In fact, comparing with eq.\ \cond, we see that
\eq
\avg\Sigma \equiv e^{i\avg{\pi_A} L_A / f} = U.
\eeq

The parameters $a$, $b_1$, and $b_2$ must be determined by matching this
effective theory onto the underlying \GUC\ gauge theory, which requires
solving a nonperturbative problem.
The parameter $a$ can be determined by using QCD as an analog computer, since
a similar term gives rise to the $\pi$ and $K$ masses.
It turns out that $a \simeq -30$, which tends to stabilize the vacuum $U = 1$.
In CTC models $K = K'$, and in order for these models to work, one requires
that the potential be minimized by
\eq
\label\Georgivac
U = \pmatrix{0 & 1\cr 1 & 0 \cr}.
\eeq
This can only be the correct vacuum if the contribution to the effective
potential from flavor gauge boson exchange stabilizes this vacuum.
This requires
\eq
g^4 \left[ b_1 - \frac{b_2}K + 3\ln 2
+ \frac 32 \ln\frac{g^2 f^2}{\mu^2} \right]
> -\frac{32\pi^2}{K^2 - 1} \; \frac af \tr m_\chi
\simeq \frac{10^3}f \tr m_{\rm_\chi}.
\eeq
If at least one of the eigenvalues of $m_\chi$ is of order $f$, as required by
the models of refs.\ \CTSM\Georgi\Kingmod, we find that even if assume
that the left-hand-side is positive, the inequality is satisfied only for
flavor gauge couplings $\alpha \gsim 5$ if we believe the NDA estimate
that $|b_1|, |b_2| \sim 1$ for $\mu \sim 4\pi f$.
This critical value for $\alpha$ is well outside the perturbative regime.
(For comparison, studies of the QCD Schwinger--Dyson equations in ladder
approximation typically show that chiral symmetry breaking occurs at a
critical coupling $\alpha_s \sim 0.8$.)
We conclude that there is no reason to expect the CTC models to choose the
vacuum of eq.\ \Georgivac.

If all of the eigenvalues of $m_\chi$ are small, then the vacuum alignment is a
more delicate question, and it is clear that a detailed calculation is
required.
This situation might arise in models similar to the CTC models in the
literature, but where the top mass arises through some other mechanism.
This would give rise to a GIM mechanism for the first two generations.
However, the computation of the next section shows that in a limit where the
vacuum alignment question can be settled with reasonable certainty, $U = 1$
is still the preferred vacuum for perturbative values of the flavor gauge
coupling.

Before presenting the formalism of the next section, it may be worthwhile to
ask
if there is a simple physical picture which suggests the answer to the vacuum
alignment question.
In the case where the effective potential induced by weak gauge boson exchange
is of order $g^2$, we can view the the effective potential as arising from the
Coulomb force between fermions.
We then expect the condensate to form in the most attractive channel, an
expectation which is born out by detailed calculations.
In the present case, there is {\it no} classical force between $\Psi_L$ and
$\Psi_R$, since they transform under independent gauge groups.
A force between $\Psi_L$ and $\Psi_R$ develops only as a result of
spontaneous symmetry breaking, and this force itself depends on the alignment.
This makes it difficult to conceive of a convincing intuitive picture, and
we must therefore rely on the more formal considerations of the next section.

\section{Computation of the Effective Potential}

We wish to evaluate the vacuum energy as a function of the orientation of the
condensate.
We will neglect the contribution from the $\chi$ mass terms and focus on the
effects of flavor gauge boson exchange.
The relevant formalism is due to many authors \vacalign.

To study the vacuum alignment question, we choose a canonical vacuum state
$\ket 0$ appropriate for the condensate \cond\ with $U = 1$.
We then consider the energy of the trial state
\eq
\ket U \equiv \hat U \ket 0,
\eeq
where $\hat U$ is the operator representing a \KLR\ transformation with
$U_L U^\dagger_R = U$ which specifies the vacuum alignment.
We can evaluate the effective potential which determines the vacuum alignment
by evaluating
\eq
V_{\rm eff}(U) = \bra U \scr H \ket U
= \bra 0 \hat U^\dagger \scr H \hat U \ket 0,
\eeq
where $\scr H$ is the hamiltonian.
The calculation is done in two steps.
First, we perform the functional integral over the \GUC\ gauge fields.
Since the \GUC\ couplings preserve the full \KKLR\ global symmetry, this
computation is independent of $U$.
We then perform the functional integral over the \KLR\ gauge fields
perturbatively, using the the Euclidean space interaction\footnote{$^\dagger$}
{We work in Euclidean space in order to simplify the job of keeping track
of signs.
Our conventions are that $A_{\mu a}$ and $J_{\mu a}$ are hermitian
operators, so that their two-point functions are positive-definite.}
\eq
\label\theint
\scr L_{\rm int} = g A_{L\mu a} J^{(U)}_{L\mu a}
+ g A_{R\mu a} J^{(U)}_{R\mu a},
\eeq
where
\eq
J^{(U)}_{L \mu a} \equiv
\mybar\Psi_L i\gamma_\mu U_L^\dagger T_{La} U_L \Psi_L,
\eeq
{\it etc\/}.
The vacuum energy is then given by {\it minus} the sum of connected Euclidean
vacuum--vacuum graphs.

The graphs which contribute to the effective potential at order $g^4$ are
shown in fig.\ 1.
At order $g^4$, the contribution of fig.\ 1b is independent of $U$.
The order $g^4$ contribution of fig 1c comes from the four-point function
$\avg{J_L J_L J_R J_R}$.
This contribution is easily seen to be subleading in the large-$N$ limit.
Thus, we need only evaluate fig.\ 1a in the large-$N$ limit.

Fig.\ 1a gives rise to the effective potential \reffpot
\eq
\label\theff
V_{\rm eff}(U) = \frac 32 \myint\frac{d^4 k}{(2\pi)^4} \,
\tr \ln \Pi^{(U)}(k^2),
\eeq
where the trace is over the adjoint representation of \KLR, and the
gauge boson vacuum polarization has been written
\eq
\Pi^{(U)}_{\mu\nu}(k) = \left( \delta_{\mu\nu} - k_\mu k_\nu / k^2 \right)
\Pi^{(U)}(k^2).
\eeq
Working to order $g^2$, we can write
\eq
\Pi^{(U)}(k^2) = \Pi_0(k^2) \cdot 1 + \delta M^2(k^2),
\eeq
where $\Pi_0$ is the diagonal and $U$-independent part of the vacuum
polarization and
\eqa
\delta M^2(k^2) & \equiv \pmatrix{0 & \delta m^2(k^2) \cr
\delta m^2(k^2) & 0 \cr}, \eol
\delta m_{ab}^2(k^2) & \equiv -g^2 \tr\left( T_{La} U T_{Rb} U^\dagger \right)
\avg{J_L(k) J_R(-k)}, \eeol
\eeq
where we have defined
\eq
\avg{J_{L\mu A}(k) J_{R\nu B}(-k)} \equiv
\left( \delta_{\mu\nu} - k_\mu k_\nu / k^2 \right) \delta_{AB}
\avg{J_L(k) J_R(-k)}.
\eeq

To estimate $\delta m^2$, we note that the operator product expansion (OPE)
tells us that
\eq
\label\ope
\avg{J_L(k) J_R(-k)} \sim \frac 1{k^4}
\quad\hbox{for $k^2 \goto \infty$.}
\eeq
This will guarantee that the $U$-dependent part of eq.\ \theff\ converges.
Imposing eq.\ \ope\ and assuming that $\avg{J_L J_R}$ is dominated by
the lowest-lying vector and axial vector resonances, we obtain \SFSR
\eq
\label\thest
\avg{J_L(k) J_R(-k)} \simeq f^2 m_\rho^2 m_a^2 \,
\frac 1{k^2 + m_\rho^2} \; \frac 1{k^2 + m_a^2}.
\eeq
Here, $\rho$ is the lowest-lying vector resonance, and $a$ is the lowest-lying
axial resonance.
We will estimate their masses by scaling up QCD.
The estimate eq.\ \thest\ is supported by $e^+$--$e^-$ annihilation
and $\tau$ decay data \PT, as well as the classic calculation
of the $\pi^+$--$\pi^0$ mass difference \pp\ and by the success of the
resulting relations among the resonance parameters \SFSR.
(See also refs.\ \vacalign.)

When we use the estimate eq.\ \thest, we see that the $U$-dependent part of
$V$ is sensitive only to $\Pi_0(k^2)$ at low momenta.
We therefore approximate
\eq
\Pi_0(k^2) \simeq (k^2 + g^2 f^2) \cdot 1.
\eeq
We expect that the corrections to this approximation become important for
momenta $k^2 \gsim m_\rho^2$, and so corrections to $V$ will be suppressed by
$\sim g^2 f^2 / m_\rho^2$.

Our task is therefore to calculate
\eq
V(U) = \frac 32 \myint \ddp 4k \tr\ln\left[ 1 + \delta M^2(k^2) \,
\frac 1{k^2 + g^2 f^2} \right],
\eeq
where we have added an irrelevant $U$-independent constant.
This may be expanded as an infinite series in $\delta M^2$.
In all but the first two terms in this series, the integral is dominated by
low momenta, and we can approximate
$\delta M^2(k^2) \simeq \delta M^2(0)$.
In this way, we find that
\eq
\label\theresult
\eqalign{
V(U) & = - \frac {3g^4 f^4}{128\pi^2} \,
\left[ 1 - \frac{2m_a^2}{m_a^2 - m_\rho^2} \ln\frac{m_a^2}{m_\rho^2} \right]
\tr\left(T_{La} U T_{Rb} U^\dagger \right)
\tr\left(T_{La} U T_{Rb} U^\dagger \right) \cr
&\qquad + \frac 3{64\pi^2} \tr M^4 \ln\frac{M^2}{m_\rho^2}. \cr}
\eeq
Comparing to the form of the low-energy effective lagrangian, we can read off
$b_1 \simeq +0.8$ for $\mu = m_\rho$.

Even in the absence of $\chi$ mass terms, eq.\ \theresult\ predicts that the
vacuum eq.\ \Georgivac\ required by CTC models occurs only for flavor gauge
couplings
$\alpha \gsim 2$.
We conclude that there is no reason to expect that the vacuum required by
CTC models will occur.

\ignore
We now consider the corrections to the large-$N$ results.
The order $g^4$ corrections are given by the diagram of fig.\ 1c, which
can be written
\eq
\label\thecorr
\eqalign{
\delta V(U) & = g^4 P_{ABCD}(U) \myint \ddp 4{k_1} \ddp 4{k_2}
\Delta_{\mu\nu}(k_1) \, \Delta_{\lambda\kappa}(k_2) \cr
&\qquad\qquad\qquad\qquad\qquad
\avg{J_{L\mu A}(k_1) J_{L\nu B}(-k_1)
J_{R\lambda C}(k_2) J_{R\kappa D}(-k_2)}, \cr}
\eeq
where $P$ is a $U$-dependent projection operator.
Neither reliable theoretical information nor QCD experimental results are
available for four-point functions such as the one in this expression.
Nonetheless, we will attempt to estimate this contribution using general
considerations.
The OPE tells us that as $k^2 \goto \infty$,
\eq
\label\opie
J_{L\mu A}(k) J_{L\nu B}(-k) \sim \delta_{\mu\nu} \delta_{AB} k^2 \cdot 1
+ \frac 1{k^4} \, \left( J_{L\mu A} J_{L\nu B} \right)(x = 0)
+ \cdots,
\eeq
{\it etc\/}.
It is easy to show that $P_{AACD}$ and $P_{ABCC}$ are independent of $U$, so
that eq.\ \opie\ guarantees that the $U$-dependent part of
eq.\ \thecorr\ converges.
Further, we can take the gauge propagators to be massless, since we are
interested only in the $O(g^4)$ contribution.
If we assume that the asymptotic $1 / k^4$ behavior predicted by the OPE
sets in for $k^2 \gsim m_\rho^2$, then we are justified in estimating
$\delta V$ by using the low-energy effective lagrangian with a momentum cutoff
of order $m_\rho^2$.
This argument justifies the NDA estimate made above, and we conclude that
$|b_2| \sim 1 / N$ in eq.\ \potform.
($\delta V$ contains contributions to both $b_1$ and $b_2$ in eq.\ \potform.)
\endignore

\section{Conclusions}

We have argued that the recently-proposed ``composite technicolor'' models
favor a phenomenologically unacceptable vacuum as long as the ``flavor''
gauge coupling can be treated perturbatively.
If the top quark gets its mass from some other mechanism, the result was
established only in the limit where the number of ultracolors $N$ is large.
While it is still possible that the $1 / N$ corrections or non-perturbative
effects for large values of the flavor gauge coupling change the preferred
vacuum, there seems to be no evidence to support such an assertion.

\section{Acknowledgements}

I would like to thank R.\ Sundrum for many noisy discussions on the subject of
this paper.
This work was supported by the Director, Office of Energy Research, Office of
High Energy and Nuclear Physics, Division of High Energy Physics of the
U.S.\ Department of Energy under Contract DE-AC03-76SF00098.


\listrefs
\vfill
\eject
\centerline{\bf Figure Captions}
\vskip .4in
\noindent
Fig.\ 1.
Contributions to the vacuum energy in the CTC model to order $g^4$.
The solid lines represent flavor gauge boson propagators, and
the shaded blobs represent the gauge boson vertex functions.

\bye